\documentclass[aps,prl,reprint,superscriptaddress]{revtex4-1}
\usepackage{graphicx}
\usepackage{color,amsmath}
\usepackage[normalem]{ulem}
\usepackage{hyperref}
\usepackage{lipsum}

\newcommand{\Vw}{V_{\rm{W}}}
\newcommand{\Vt}{V_{\rm{t}}}
\newcommand{\Bp}{B_{\parallel}}
\newcommand{\Br}{B_{\rm{r}}}
\newcommand{\Vsd}{V_{\rm{SD}}}
\newcommand{\Gp}{G_{\rm{P}}}
\newcommand{\Gn}{G_{\rm{N}}}
\newcommand{\Go}{\times2e^2/h}
\newcommand{\Dt}{\Delta_{\rm{t}}}

\begin{document}

\title{Scaling of Majorana Zero-Bias Conductance Peaks}

\author{Fabrizio~Nichele}
\email[email: ]{fnichele@nbi.ku.dk}
\affiliation{Center for Quantum Devices and Station Q Copenhagen, Niels Bohr Institute, University of Copenhagen, Universitetsparken 5, 2100 Copenhagen, Denmark}

\author{Asbj\o rn~C.~C.~Drachmann}
\affiliation{Center for Quantum Devices and Station Q Copenhagen, Niels Bohr Institute, University of Copenhagen, Universitetsparken 5, 2100 Copenhagen, Denmark}

\author{Alexander~M.~Whiticar}
\affiliation{Center for Quantum Devices and Station Q Copenhagen, Niels Bohr Institute, University of Copenhagen, Universitetsparken 5, 2100 Copenhagen, Denmark}

\author{Eoin~C.~T.~O'Farrell}
\affiliation{Center for Quantum Devices and Station Q Copenhagen, Niels Bohr Institute, University of Copenhagen, Universitetsparken 5, 2100 Copenhagen, Denmark}

\author{Henri~J.~Suominen}
\affiliation{Center for Quantum Devices and Station Q Copenhagen, Niels Bohr Institute, University of Copenhagen, Universitetsparken 5, 2100 Copenhagen, Denmark}

\author{Antonio~Fornieri}
\affiliation{Center for Quantum Devices and Station Q Copenhagen, Niels Bohr Institute, University of Copenhagen, Universitetsparken 5, 2100 Copenhagen, Denmark}

\author{Tian~Wang}
\affiliation{Department of Physics and Astronomy and Station Q Purdue, Purdue University, West Lafayette, Indiana 47907 USA}
\affiliation{Birck Nanotechnology Center, Purdue University, West Lafayette, Indiana 47907 USA}

\author{Geoffrey~C.~Gardner}
\affiliation{School of Materials Engineering, Purdue University, West Lafayette, Indiana 47907 USA}
\affiliation{Birck Nanotechnology Center, Purdue University, West Lafayette, Indiana 47907 USA}

\author{Candice~Thomas}
\affiliation{Department of Physics and Astronomy and Station Q Purdue, Purdue University, West Lafayette, Indiana 47907 USA}
\affiliation{Birck Nanotechnology Center, Purdue University, West Lafayette, Indiana 47907 USA}

\author{Anthony~T.~Hatke}
\affiliation{Department of Physics and Astronomy and Station Q Purdue, Purdue University, West Lafayette, Indiana 47907 USA}
\affiliation{Birck Nanotechnology Center, Purdue University, West Lafayette, Indiana 47907 USA}

\author{Peter~Krogstrup}
\affiliation{Center for Quantum Devices and Station Q Copenhagen, Niels Bohr Institute, University of Copenhagen, Universitetsparken 5, 2100 Copenhagen, Denmark}

\author{Michael~J.~Manfra}
\affiliation{Department of Physics and Astronomy and Station Q Purdue, Purdue University, West Lafayette, Indiana 47907 USA}
\affiliation{School of Materials Engineering, Purdue University, West Lafayette, Indiana 47907 USA}
\affiliation{School of Electrical and Computer Engineering, Purdue University, West Lafayette, Indiana 47907 USA}
\affiliation{Birck Nanotechnology Center, Purdue University, West Lafayette, Indiana 47907 USA}

\author{Karsten~Flensberg}
\affiliation{Center for Quantum Devices and Station Q Copenhagen, Niels Bohr Institute, University of Copenhagen, Universitetsparken 5, 2100 Copenhagen, Denmark}

\author{Charles~M.~Marcus}
\affiliation{Center for Quantum Devices and Station Q Copenhagen, Niels Bohr Institute, University of Copenhagen, Universitetsparken 5, 2100 Copenhagen, Denmark}

\date{\today}

\begin{abstract}
We report an experimental study of the scaling of zero-bias conductance peaks compatible with Majorana zero modes as a function of magnetic field, tunnel coupling, and temperature in one-dimensional structures fabricated from an epitaxial semiconductor-superconductor heterostructure. Results are consistent with theory, including a peak conductance that is proportional to tunnel coupling, saturates at $2e^2/h$, decreases as expected with field-dependent gap, and collapses onto a simple scaling function in the dimensionless ratio of temperature and tunnel coupling.
\end{abstract}

\maketitle

\begin{figure*}
\includegraphics[width=2\columnwidth]{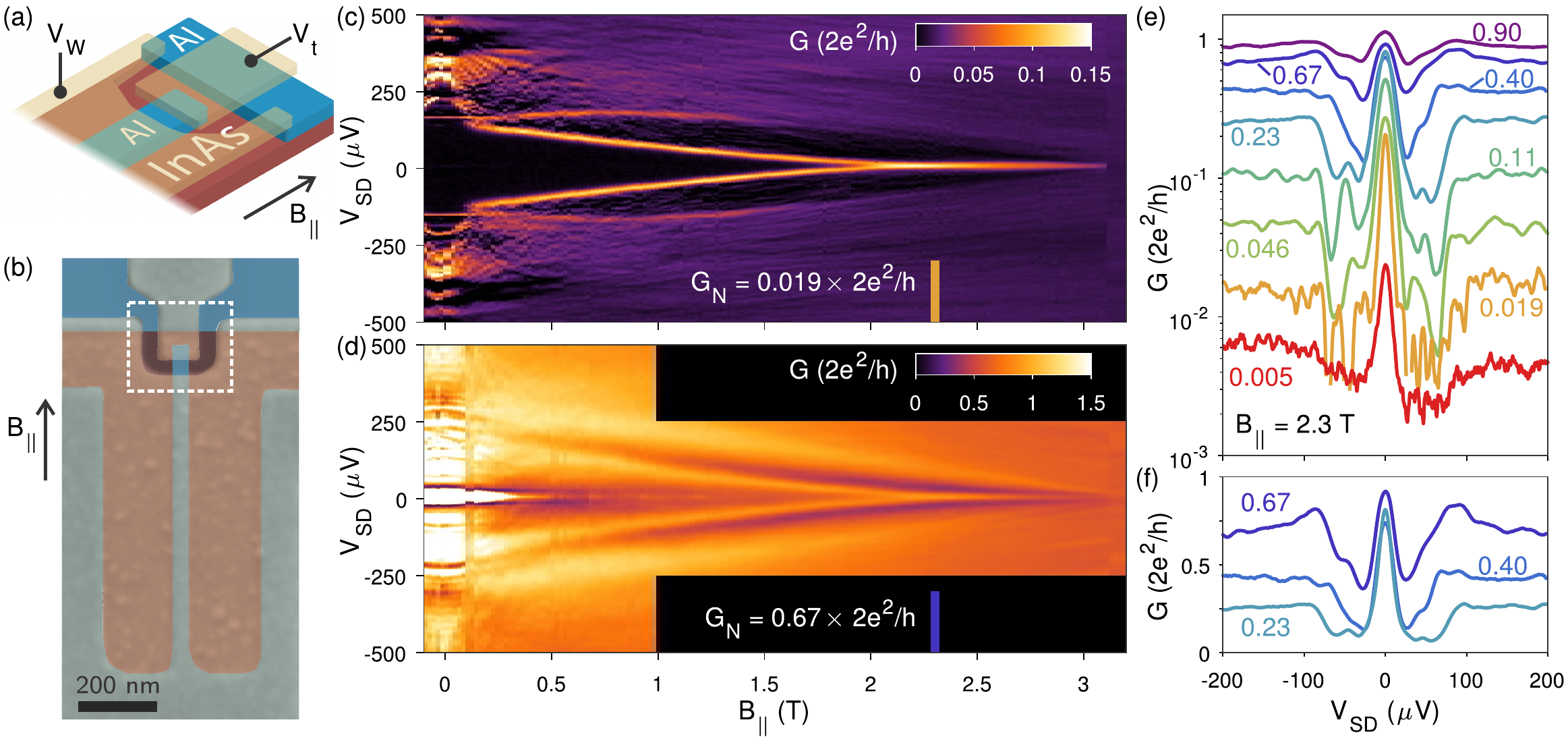}
\caption{(a) Device schematic close to the tunneling junction, as in the dashed box in (b). The end of the Al wire and the top Al plane (blue) are separated by a narrow InAs junction (red). Two gates (yellow) allow independent tuning of the chemical potential in the wire (gate voltage $\Vw$) and of the junction transmission (gate voltage $\Vt$). (b) False color scanning electron micrograph of a typical device. The Al pattern is visible through the top gate. The epitaxial Al below the gates is colored gray, the semiconductor below the gates is colored red. (c) Tunneling conductance as a function of in-plane magnetic field $\Bp$ for normal state transmission $\Gn=0.019\Go$. The colored line indicates the position where the linecuts of (e) and (f) were taken. Color extrema have been saturated. (d) As in (c) for $\Gn=0.67\Go$. (e) ZBP conductance at $\Bp=2.3~\rm{T}$ for several values of $\Gn$, indicated in the figure in units of $2e^2/h$. (f) Selection of curves from (e), plotted on a linear scale.}
\label{fig1}
\end{figure*}

Recent years have seen rapid progress in the study of Majorana zero modes (MZMs) in condensed matter. Following initial reports of zero-bias peaks (ZBPs) in conductance of nanowire-superconductor hybrids appearing at moderate magnetic fields \cite{Mourik2012}, improvements in materials \cite{Krogstrup2015,Chang2015,Guel2017} resulted in harder induced gaps and the emergence of zero-bias peaks from coalescing Andreev bound states (ABSs) \cite{Zhang2016,Deng2016}, as well as the observation of exponential suppression of Coulomb peak oscillations with nanowire length \cite{Albrecht2016}. Recently, indications of MZMs were also identified in wires lithographically patterned on hybrid two-dimensional heterostructures \cite{Shabani2015,Suominen2017}. In many respects, experimentally observed ZBPs are consistent with theoretical expectations for MZMs, but important questions remain, particularly concerning theoretical models that show ZBPs arising from nontopological ABSs in localized states at the wire ends \cite{Lee2013,Liu2017a}. Furthermore, the fact that observed zero-bias peaks \cite{Mourik2012,Das2012,Zhang2016,Deng2016} were considerably smaller than the theoretically expected value of $2e^2/h$ \cite{Sengupta2001,Law2009,Akhmerov2009,Flensberg2010,Wimmer2011,Zazunov2016} has raised concern. Speculations about the origin of this discrepancy included effects of dissipation \cite{Liu2017} as well as nontopological ZBPs induced by disorder \cite{Liu2012,Bagrets2012,Pikulin2012} or a spin-orbit-induced precursor \cite{Lee2013}.

In this Letter, we investigate ZBPs in lithographically defined wires as a function of temperature, tunnel coupling to a metallic lead (parametrized by the normal state conductance $\Gn$), and magnetic field. For weak coupling to the lead ($\Gn \ll e^{2}/h$), a small ZBP with strong temperature dependence is observed over an extended range of magnetic fields. For strong coupling ($\Gn \sim e^{2}/h$), the dependence of the ZBP on $\Gn$ and temperature weakens, with a low-temperature saturation at $\sim2e^2/h$. Experimental results are well described by a theoretical model of resonant transport through a zero-energy state that includes both broadening due to coupling to a normal lead and temperature.

Fitting ZBP heights as a function of temperature, $T$, and $\Gn$ yields values for the energy broadening, $\Gamma$, which we find obey the linear relationship $\Gamma \propto \Gn$.
The fit results for $\Gamma$ are found to be in excellent agreement with a scaling function that depends only on the dimensionless ratio $\Gamma/k_{\rm{B}}T$. The observed magnetic field dependence of the ZBP is quantitatively consistent with a picture in which field reduces the induced superconducting gap, $\Delta^*$, which in turn reduces the ZBP height through the dependence of $\Gamma$ on $\Delta^*$.

Overall, the ZBPs reported here are compatible with MZMs, and suggest that the temperature-to-broadening ratio is the limiting factor controlling ZBP conductance in setups where tunnel coupling is weak \cite{Mourik2012,Deng2016}. We emphasize that although the observed conductance saturation, scaling, and field dependence are all consistent with a MZM interpretation, these results could be obtained with specially tuned nontopological Andreev states that happen to stick to zero energy \cite{Liu2017a}. Distinguishing topological from trivial scenarios relies on examining the stability of ZBPs to tuning parameters, as discussed below.

Measurements are performed on wires lithographically defined on an InAs/Al heterostructure \cite{Shabani2015}, using the same approach as in Ref.~\onlinecite{Suominen2017}. Figures~\ref{fig1}(a) and \ref{fig1}(b) show a schematic and an electron micrograph, respectively, of a typical sample. A $1.5~\rm{\mu m}$ long, $120~\rm{nm}$ wide Al strip is defined on the wafer surface by selective Al etching. The Al strip is connected at one end with a large planar Al region, and on the other end terminates with a $40~\rm{nm}$ break separating it from another planar Al region. A HfO$_{2}$ insulating layer is deposited by atomic layer deposition over the entire sample, followed by two patterned gates separately covering the wire and the break. Applying negative voltage $\Vw$ on the gate covering the wire depletes the surrounding two-dimensional electron gas region, leaving a narrow undepleted region of the InAs quantum well screened by the Al strip. The tunnel barrier is independently controlled with the voltage $\Vt$ on the break region. Transport measurements are performed using standard low frequency lock-in techniques in a dilution refrigerator with base electron temperature of $\sim40~\rm{mK}$. Throughout this paper, we characterize the low-temperature transport measurements in terms of the normal state conductance, $\Gn$, measured as the differential conductance at large source drain bias $\Vsd$. These values coincide with the $\Vsd=0$ conductances measured above the critical temperature of the Al film. Further details of wafer structure, sample fabrication, as well as additional measurements are given in the Supplemental Material \cite{Supplement}.

For magnetic fields $\Bp$ (oriented along the wire) lower than $\sim100~\rm{mT}$, both the strip and plane regions of the quantum well covered by Al show a hard induced superconducting gap, resulting in a superconductor-insulator-superconductor (SIS) junction. At larger fields, the superconducting gap below the Al plane softens, creating a finite density of state at zero energy, while the gap below the Al strip remains hard up to $3~\rm{T}$ \cite{Suominen2017}. This feature allows the Al plane to be used, at moderate $\Bp$, as an effective normal lead, making a superconductor-insulator-normal (SIN) junction, which can be used to perform tunneling spectroscopy of the wire.

Tunneling conductance as a function of $\Bp$ at small and large barrier transmissions (controlled by $\Vt$) for similar wire densities (controlled by $\Vw$) are shown in Figs.~\ref{fig1}(c) and ~\ref{fig1}(d). For large transmission and low field, the SIS configuration is evident from a conductance enhancement at $\Vsd=0$ [up to $100\times2e^2/h$ in the data of Fig.~\ref{fig1}(d)] as well as tunneling conductance peaks at $\Vsd = \pm 2\Delta^*/e\sim \pm 400~\rm{\mu V}$ and fractions reflecting multiple Andreev reflection \cite{Blonder1982,Kjaergaard2017}. The zero bias supercurrent disappears as the transmission is lowered, while low-order multiple Andreev reflections remain. In Fig.~\ref{fig1}(d), for example, the first order Andreev reflection is visible as a conductance peak at $\Vsd=\pm\Delta^*/e\sim\pm200~\rm{\mu V}$. Regardless of transmission, the planar lead acquires effectively normal behavior above $\Bp\sim300~\rm{mT}$, as seen from the crossover of tunneling features from $\pm2\Delta^*/e$ (expected for SIS) to $\pm \Delta^*/e$ (expected for SIN).

At larger fields, $\Bp\sim2~\rm{T}$, a robust ZBP forms from subgap states moving toward zero energy. While the overall appearance of data in Fig.~\ref{fig1}(c) suggests a MZM interpretation, an explanation in terms of localized Andreev bound states that happen to stick to zero energy \cite{Liu2017a} cannot be ruled out by these measurements alone. Further support for a MZM interpretation includes field-angle and gate dependence (see Figs.~\ref{figs1} and~\ref{figs2} in Ref.~\onlinecite{Supplement}), rehardening of the gap at high field (see Fig.~\ref{figs3} in Ref.~\onlinecite{Supplement}), as well as stability of the ZBP with respect to gate voltages applied close to the lithographic end of the wire, as discussed in reference to Fig.~\ref{fig4}.

The same ZBP was measured for tunnel couplings ranging from $\Gn=0.005\times 2e^{2}/h$ to $1\times 2e^{2}/h$, by tuning $\Vt$. For low transmission, a sharp ZBP was observed on a hard gap, as seen in Fig.~\ref{fig1}(c). The peak height, $\Gp$, defined as the conductance at $\Vsd=0$ without background subtraction, was up to $10$ times higher than $\Gn$, though still considerably lower than $2e^2/h$. Increasing $\Gn$ by adjusting $\Vt$ broadened the peak and increased $\Gp$ toward $\sim 2e^2/h$, while decreasing the ratio $\Gp/\Gn$ toward 1. The linear vertical scale in Fig.~\ref{fig1}(f) emphasizes that as $\Gn$ increased by a factor of nearly 3, from $0.23$ to $0.67$ in units of $2e^2/h$, the ZBP height only changed by about $10\%$. Additionally, $\Gp$ is found to decrease as $\Bp$ increases, an effect we attribute to the collapsing gap, as discussed below.

Increasing the tunnel coupling not only increased $\Gp$, but also increased the background subgap conductance, as expected theoretically \cite{Blonder1982,Beenakker1992}. Large magnetic fields further soften the superconducting gap \cite{Chang2015,Kjaergaard2016}. For $\Gn=0.23\Go$, the zero bias conductance peak is $\Gp=0.81\times2e^2/h$, $12$ times higher than the conductance minimum at finite $\Vsd$, below the gap edge [see Figs.~\ref{fig1}(e) and ~\ref{fig1}(f)]. As the barrier transmission is further increased, the contribution from the ZBP is no longer well separated from the above-gap conductance. The analysis of the dependence of $\Gp$ on transmission presented below suggests that more than one channel participates in the transport. This presumably also accounts for the low-temperature saturation of $\Gp$ for the most open barrier exceeding $2e^2/h$ by $\sim$13\%. A ZBP is not visible for $\Gn$ above $2e^2/h$.

\begin{figure}[t]
\includegraphics[width=\columnwidth]{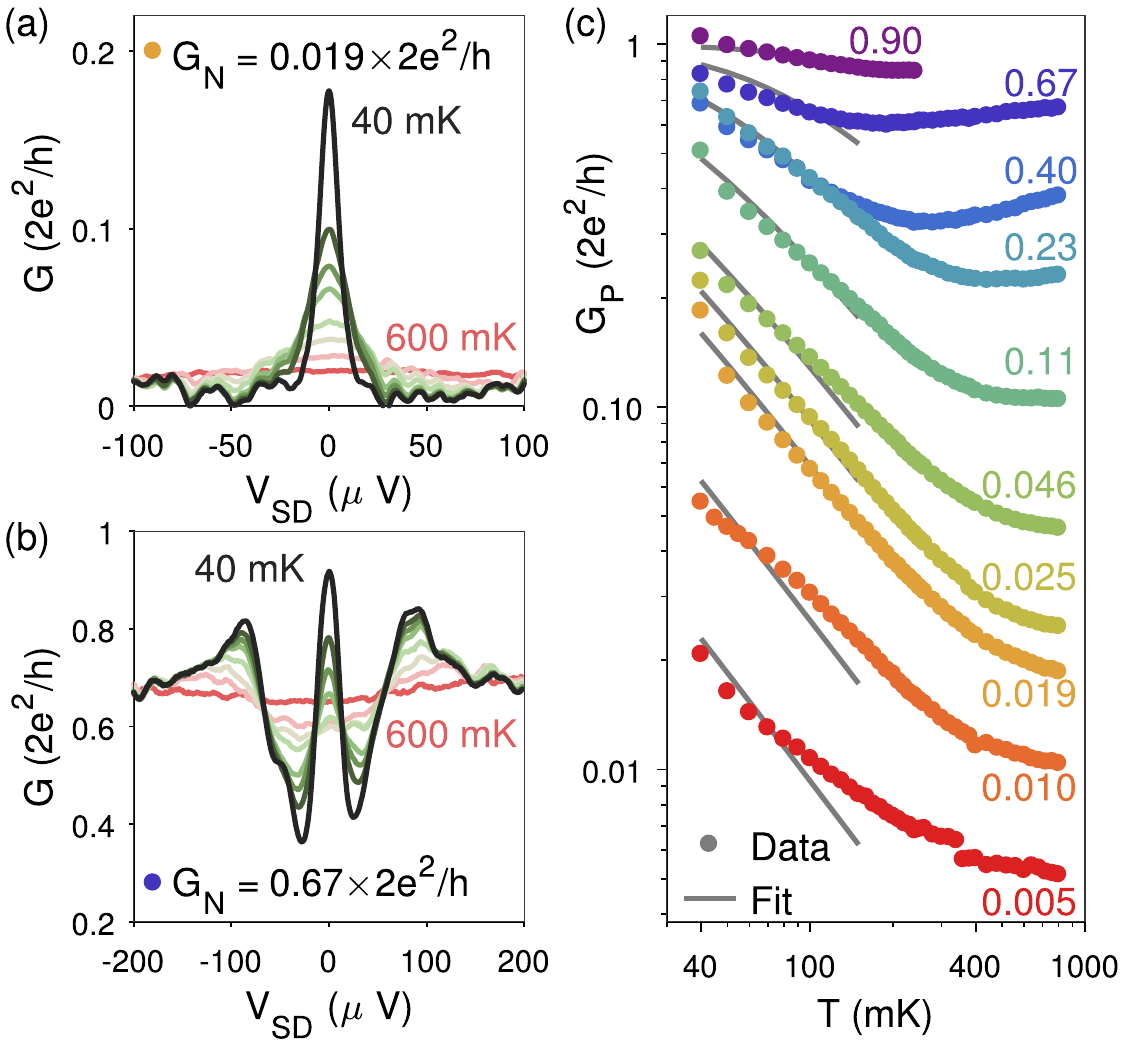}
\caption{(a) Conductance of a ZBP at $\Bp =2.3~\rm{T}$ and $\Gn=0.19\Go$ for various temperatures. Temperatures are $40$ (black), $60$, $80$, $100$, $150$, $200$, $300$, and $600~\rm{mK}$ (pink). (b) As in (a), but for $\Gn=0.67\Go$. (c) Temperature dependence ZBP conductance, $\Gp$, at $\Bp=2.3~\rm{T}$, measured for different values of $\Gn$ (colored dots) together with fits to Eq.~\eqref{Glowtemp} (gray curves). For the experimental data, the same colors as in Fig.~\ref{fig1}(e) are used.}
\label{fig2}
\end{figure}

Turning next to the temperature dependence of ZBPs at different tunnel couplings, Figs.~\ref{fig2}(a) and~\ref{fig2}(b) show cuts from Figs.~\ref{fig1}(c) and~\ref{fig1}(d) at $\Bp=2.3~\rm{T}$, for temperatures ranging from $40$ to $600~\rm{mK}$. In both cases, reducing the temperature results in an increase of the ZBP height and a reduction of its width. The temperature dependence of the ZBP height depends on transmission: lowering $T$ from $60$ to $40~\rm{mK}$ increases $\Gp$ at low transmission by $90\%$ [Fig.~\ref{fig2}(a)], while $\Gp$ at high transmission increases by only $15\%$ [Fig.~\ref{fig2}(b)]. Temperature dependence across a broad range of transmissions is shown in Fig.~\ref{fig2}(c). Consistent with the examples in Figs.~\ref{fig2}(a) and~\ref{fig2}(b), values of $\Gp$ in Fig.~\ref{fig2}(c) depend only weakly on $T$ for high transmission, near $\Gp\sim2e^2/h$, while for low transmission, $\Gp$ values depend strongly on temperature.

\begin{figure}[b]
\includegraphics[width=\columnwidth]{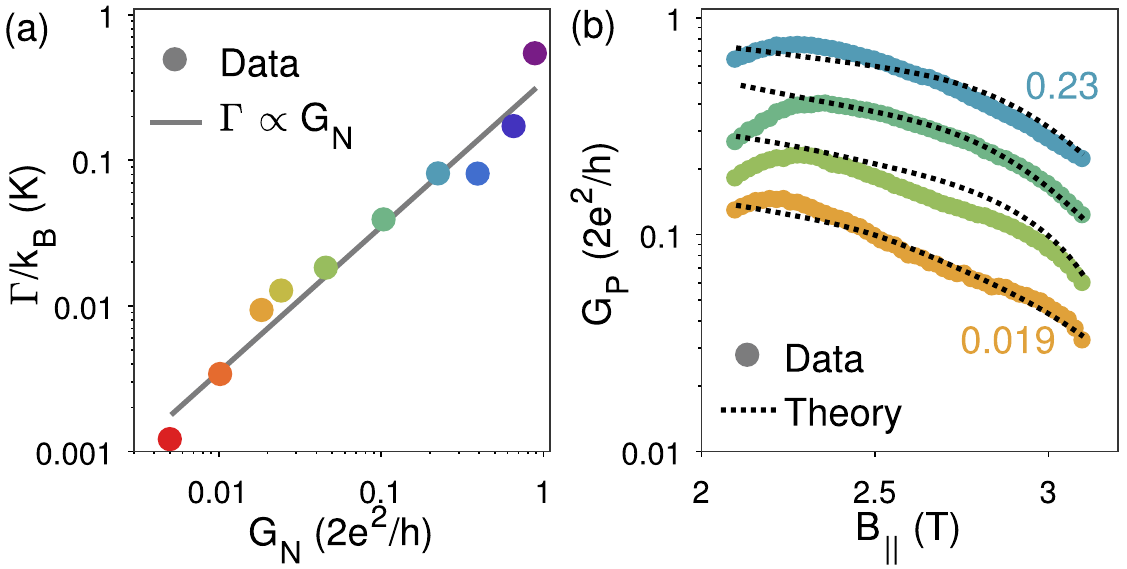}
\caption{(a) Extracted relation between the tunnel broadening, $\Gamma$, obtained from the fits in Fig.~\ref{fig2}(c), and the normal-state conductance, $\Gn$ ( dots), along with a linear fit (gray line). (b) Magnetic field dependence of the ZBP conductance, $\Gp$, together with theoretical predictions based on Eq.~(1), with $\Gamma \propto \Delta^{*}(\Bp)$.}
\label{fig3}
\end{figure}

These observations can be compared to a model of a MZM broadened by tunnel coupling and temperature. For energies below the topological gap $\Dt$, tunnel broadening gives a Lorentzian lineshape to the zero-temperature Majorana peak \cite{Sengupta2001,Flensberg2010,Zazunov2016,VanHeck2016}. For weak tunneling, the width, $\Gamma$, of the ZBP is predicted to be directly proportional to the transparency $\tau$ of the junction and the topological gap, $\Gamma\propto\tau\Dt$ \cite{Sengupta2001,Zazunov2016,VanHeck2016}. Physically, the proportionality to the gap reflects the tighter localization of the MZM to the boundary of the topological region with larger gap. According to this picture, for $k_{\rm{B}}T\ll\Dt$ the zero-bias conductance is given by
\begin{equation}\label{Glowtemp}
\begin{aligned}
 \Gp&\approx \frac{e^2}{h}\int_{-\infty}^{\infty}d\omega\frac{2\Gamma^2}{\omega^2+\Gamma^2}\frac{1}{4k_{\rm{B}}T\cosh^2(\omega/(2k_{\rm{B}}T))}\\
 &=\frac{2e^2}{h}f(k_{\rm{B}}T/\Gamma).
\end{aligned}
\end{equation}
Note that the scaling function $f$ depends only on the ratio of temperature to tunnel broadening. Fits of Eq.~\eqref{Glowtemp} to the $T\leq150~\rm{mK}$ data in Fig.~\ref{fig2}(c) yield values for the single fit parameter $\Gamma$ for each value of $\Gn$. The resulting values are shown in Fig.~\ref{fig3}(a). The fit values of $\Gamma$ were found to be proportional to $\Gn$ [gray line in Fig.~\ref{fig3}(a)] over 2 orders of magnitude. A power-law fit of the form $\Gamma\propto(\Gn)^\alpha$ yields $\alpha=1.0\pm0.1$.

A spinless model for MZM transport yields the relation $\Gamma=\tau\Dt/(2\sqrt{1-\tau})$ across the full range of transmission $0<\tau<1$ \cite{Kashiwaya2000,Sengupta2001}. On the other hand, experimentally, we find that the linear relation $\Gamma\propto\Gn$ holds even as $\Gn\rightarrow2e^2/h$ [Fig.~\ref{fig3}(a)], apparently inconsistent with the spinless model if one makes the identification $\tau=\Gn/(2e^2/h)$. We note, however, that the spinless model is not valid for $\Gamma$ of order the gap. A spinful single-mode model with transmission close to unity predicts an in-gap conductance near $4e^2/h$ at finite bias, with a dip rather than a peak at zero bias to $2e^2/h$ \cite{Wimmer2011,Setiawan2015}, which we do not observe. Conductance doubling due to Andreev reflection was recently observed by us in a different device geometry \cite{Kjaergaard2016}, but only at low field and zero bias. A hint to this behavior might be visible in the highest transmission curves of Fig.~\ref{fig1}(e), where the finite bias in-gap conductance quickly raises as transmission approaches unity.
These discrepancies between model predictions and experiment could also suggest that more transverse modes below the Al strip contribute to transport. A multimode scenario could also explain the observation of $\Gp$ exceeding $2e^2/h$ when $\Gn\sim 2e^2/h$, as seen in Fig.~\ref{fig2}(c) (violet dots) while still being in agreement with the small conductance traces, keeping in mind that the contribution to Andreev reflection from the additional modes would not significantly modify the low temperature quantization for $\tau\ll1$ \cite{Blonder1982,Beenakker1992}. These observations motivate a more detailed understanding of the finite-bias transport in multimode topological wires.

The linear fit in Fig.~\ref{fig3}(a) gives $\Gamma/k_{\rm{B}}\sim \Gn/(2e^2/h) \times 430~\rm{mK}$. With $\tau=\Gn/(2e^2/h)$, the model relation $\Gamma\approx\tau\Dt/2$ (valid for $\tau\ll 1$) yields $\Dt\sim75~\rm{\mu eV}$, which is comparable to the gap measured directly from Figs.~\ref{fig1}(c) and ~\ref{fig1}(f). The proportionality $\Gamma\propto\Dt$ suggests a mechanism for the observed reduction of the ZBP as $\Bp$ increases: quenching of $\Dt$ by the external field reduces $\Gamma$, which in turn lowers $\Gp$ through the ratio $k_{B}T/\Gamma$ [see Eq.~\eqref{Glowtemp} and Fig.~\ref{fig4}]. To test this connection quantitatively, we used $\Delta^*(\Bp)$ from various measurements including data in Figs.~\ref{fig1}(c) and ~\ref{fig1}(d) to calculate $\Gp$, assuming $\Gamma(\Bp)=\tau\Delta^*(\Bp)/2$ and Eq.~\eqref{Glowtemp}. Figure~\ref{fig3}(b) shows good agreement between experiment and this simple calculation, indicating that the reduction of the ZBP at high field is a consequence of the closing of the induced gap \footnote{The theory curves are calculated for $T=55~\rm{mK}$ due to the elevated electron temperature during magnetic field sweeps \cite{Supplement}}. We note that the observed decrease in ZBP height with decreasing $\Delta^{*}(\Bp)$ suggests a mode fixed to the end of the wire. In contrast, for a mode away from the end of the wire (on the scale of a coherence length) an increase in tunnel coupling as $\Delta^{*}(\Bp)$ decreases would be expected.

\begin{figure}
\includegraphics[width=\columnwidth]{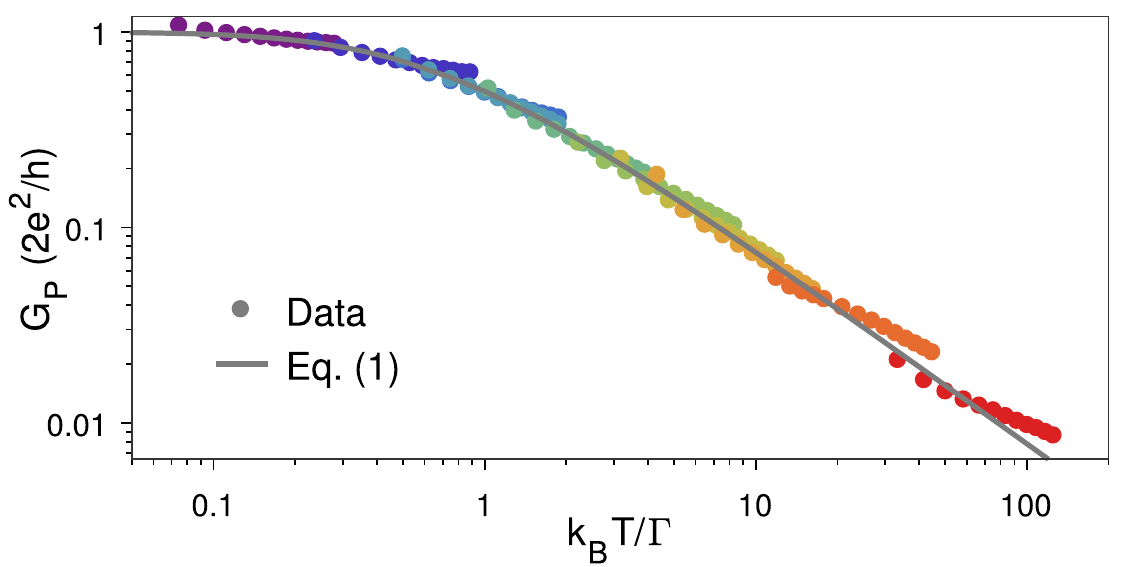}
\caption{Zero bias conductance of the entire data set (for $T\leq150~\rm{mK}$) plotted as a function of temperature scaled by the extracted values of $\Gamma$ (colored dots) and compared to the theoretical expectation of Eq.~\eqref{Glowtemp} (gray curve).}
\label{fig4}
\end{figure}

Zero-bias peak conductances for the full range of the dimensionless scale factor $k_{\rm{B}}T/\Gamma$ are shown in Fig.~\ref{fig4} along with the scaling function, Eq.~\eqref{Glowtemp}, for transport through a single zero-energy mode. A striking consistency over 3 orders of magnitude in $k_{\rm{B}}T/\Gamma$, with low-temperature saturation $2e^2/h$ is observed. We emphasize that scaling and saturation at $2e^2/h$, as expected for a MZM, does not rule out a nontopological discrete state at zero energy, for instance of the type discussed in Ref.~\onlinecite{Liu2017a}, as the origin of the ZBP. Support for a MZM interpretation includes the robustness of features to variation in $\Bp$, $\Vw$, and $\Vt$. We found that gate voltage $\Vt$ tunes the tunneling probe transmission up to 3 orders of magnitude but does not affect the ZBP except for the behavior captured by Eq.~\eqref{Glowtemp}. A disorder-induced bound state near the end of our wires would presumably be affected by $\Vt$ [see Fig.~\ref{fig1}(a)], resulting in variations of occupation, low temperature conductance, and behavior in a magnetic field. Essentially identical behavior seen in a second device \cite{Supplement} further suggests a MZM rather than a localized state resulting from disorder.

A similar single parameter scaling has been applied to Kondo resonances in quantum dots \cite{Wiel2000}, including devices with superconducting leads \cite{Lee2012}. Qualitatively similar to Fig.~\ref{fig4}, the Kondo resonance results in a ZBP with low temperature saturation to $2e^2/h$ and a suppression at high temperature, despite with a different functional form than Eq.~\eqref{Glowtemp}. The emergence of the ZBP from Andreev states converging at zero energy disfavors a Kondo interpretation. Also, the Kondo resonance typically requires low magnetic fields and symmetric leads. Here we operate in a very large field and with a single tunnel barrier whose transmission is tuned by more than 2 orders of magnitude without affecting the presence of the ZBP.

In conclusion, we have investigated ZBPs in a Majorana device patterned in a two-dimensional heterostructure with epitaxial Al. The devices design allows a systematic study of the conductance for different values of the tunnel broadening. The low-temperature data show a scaling behavior where the peak height follows a simple universal curve that depends only on the dimensionless parameter $k_{\rm{B}}T/\Gamma$ and saturates at $2e^2/h$ for $k_{\rm{B}}T/\Gamma\ll1$. These results suggest that small ZBPs previously reported may be compatible with MZMs if they were obtained in a regime where the ratio of temperature to broadening was large.

\begin{acknowledgments}
This work was supported by Microsoft Corporation, the Danish National Research Foundation, the Villum Foundation, and the DFG Mercator program.
\end{acknowledgments}

\bibliography{Bibliography}

\setcounter{figure}{0}
\renewcommand{\thefigure}{S.\arabic{figure}}
\renewcommand{\theHfigure}{Supplement.\thefigure}

\newpage

\section{Material and Methods}
The wafer structure used in this work was grown by molecular beam epitaxy on a semi-insulating InP substrate. From the bottom to top it consisted of an $100~\rm{nm}$ $\rm{In_{0.52}Al_{0.48}As}$ buffer, a 5-period $2.5~\rm{nm}$ $\rm{In_{0.53}Ga_{0.47}As/2.5~nm~In_{0.52}Al_{0.48}As}$ superlattice, a $1~\rm{\mu m}$ thick metamorphic graded buffer stepped from $\rm{In_{0.52}Al_{0.48}As}$ to $\rm{In_{0.84}Al_{0.16}As}$, a $33~\rm{nm}$ graded $\rm{In_{0.84}Al_{0.16}As}$ to $\rm{In_{0.81}Al_{0.19}As}$ layer, a $25~\rm{nm}$ $\rm{In_{0.81}Al_{0.19}As}$ layer, a $4~\rm{nm}$ $\rm{In_{0.81}Ga_{0.19}As}$ lower barrier, a $5~\rm{nm}$ InAs quantum well, a $10~\rm{nm}$ $\rm{In_{0.81}Ga_{0.19}As}$ top barrier, two monolayers of GaAs and finally a $8.7~\rm{nm}$ layer of epitaxial Al.
The top Al layer has been grown in the same molecular beam epitaxy chamber used for the rest of the growth, without breaking the vacuum. The epitaxial growth results in semiconductor/superconductor interfaces characterized by almost unitary transparency \cite{Kjaergaard2017}. The two monolayers of GaAs are introduced to help passivate the wafer surface where the Al film is removed, and to make the sample more compatible with our Al etchant (see below), which does not attack GaAs. The two-dimensional electron gas (2DEG) is expected to mainly reside in the InAs quantum well, with the upper tail of the wavefunction extending to the Al film \cite{Shabani2015}.

Characterization performed in a Hall bar geometry where the Al was removed revealed an electron mobility peak $\mu=18000~\rm{cm^2V^{-1}s^{-1}}$ for an electron density $n=9\times10^{11}~\rm{cm^{-2}}$. Characterization of a large area Al film revealed a critical magnetic field $B_{\rm{c}}=2.85~\rm{T}$ when the field is applied in the plane of the 2DEG. On the other hand, fine Al structures, such as the wires measured in our devices, exhibit $B_{\rm{c}}\approx3.1~\rm{T}$.

\begin{figure}
\includegraphics[width=\columnwidth]{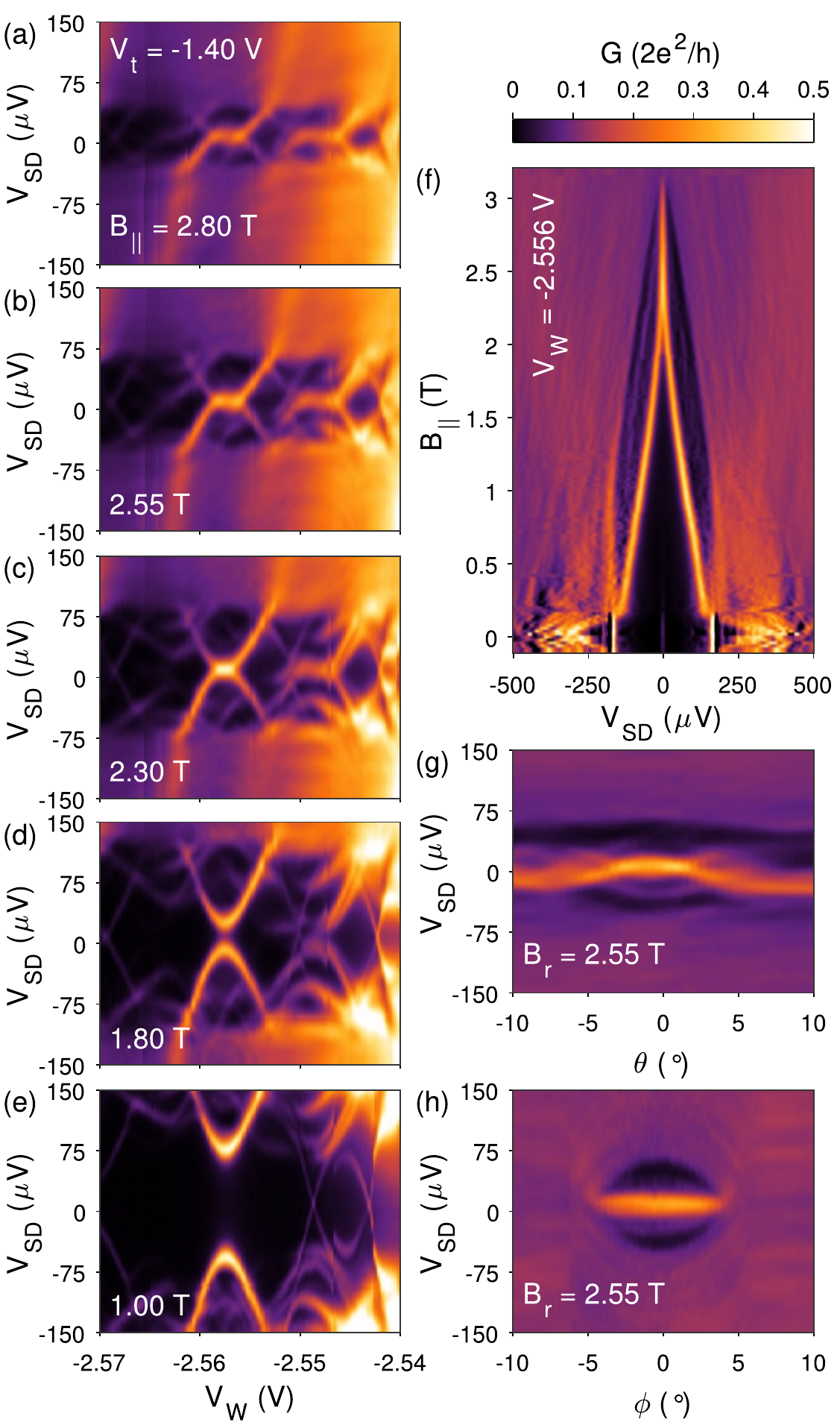}
\caption{(a-e) Bias spectroscopy of the same ZBP as a function of $\Vw$ for decreasing values of $\Bp$, reported in each subfigure, and for $\Vt=-1.40~\rm{V}$. (f) Bias spectroscopy of the ZBP shown in (a-e) for $\Vw=-2.555~\rm{V}$ as a function of $\Bp$. (g) ZBP conduction as a function of magnetic field orientation $\theta$ for a fixed magnetic field magnitude $\Br=2.55~\rm{T}$. $\theta$ gives the angle on the plane of the 2DEG, with $\theta=0$ coinciding with the wire direction. (h) Same as (g), but for the magnetic field aligned at an angle $\phi$ that lies on the plane perpendicular to the 2DEG and parallel to the wire. All the subfigures share the same colorbar, saturated to $0.5\times2e^2/h$.}
\label{figs1}
\end{figure}

Samples were fabricated with conventional electron beam lithography techniques. First, we isolated large mesa structures by locally removing the top Al layer (Al etchant Transene D) and performing a deep III-V chemical wet etch (220:55:3:3 $\rm{H_2 O:C_6 H_8 O_7:H_3 PO_4:H_2 O_2}$). In a subsequent step, we patterned the Al wires by selectively removing the top Al layer with a wet etch (Al etchant Transene D). All the Al etching steps are performed at a temperature of $50^\circ\rm{C}\pm1^\circ\rm{C}$ for $5~\rm{s}$.
We then deposit, on the entire sample, a $40~\rm{nm}$ thick layer of high-$\kappa$ dielectric $\rm{HfO_2}$ by atomic layer deposition at a temperature of $90^\circ\rm{C}$. The top gate electrodes are deposited in two successive steps. First we define the features requiring high accuracy and deposit $5~\rm{nm}$ of Ti and $25~\rm{nm}$ of Au by electron beam evaporation. In a successive step, we define the gates bonding pads by evaporating $10~\rm{nm}$ Ti and $250~\rm{nm}$ Au. Ohmic contacts to the InAs are provided by the epitaxial Al layer, which is contacted directly by wedge bonding through the insulating $\rm{HfO_2}$. As explained in the Main Text, with reference to Fig.~\ref{fig1}(a) and~\ref{fig1}(b), the sample was tuned by the gate voltages $\Vw$ and $\Vt$ respectively. The gate voltage $\Vw$ mainly influences the chemical potential in the InAs channel, below the Al strip, while $\Vt$ tunes the coupling between the wire and the lead. 
Two lithographically similar samples were studied, named Sample~1 and Sample~2, fabricated on a different chip obtained from the same wafer structure. Both devices had a wire length of $1.5~\rm{\mu m}$. Sample~2 had a wire width of $100~\rm{nm}$ whereas it was $120~\rm{nm}$ in Sample~1. Data in the Main Text exclusively refers to Sample~1. Data in this Supplemental Material reefers to both samples.

\begin{figure}
\includegraphics[width=\columnwidth]{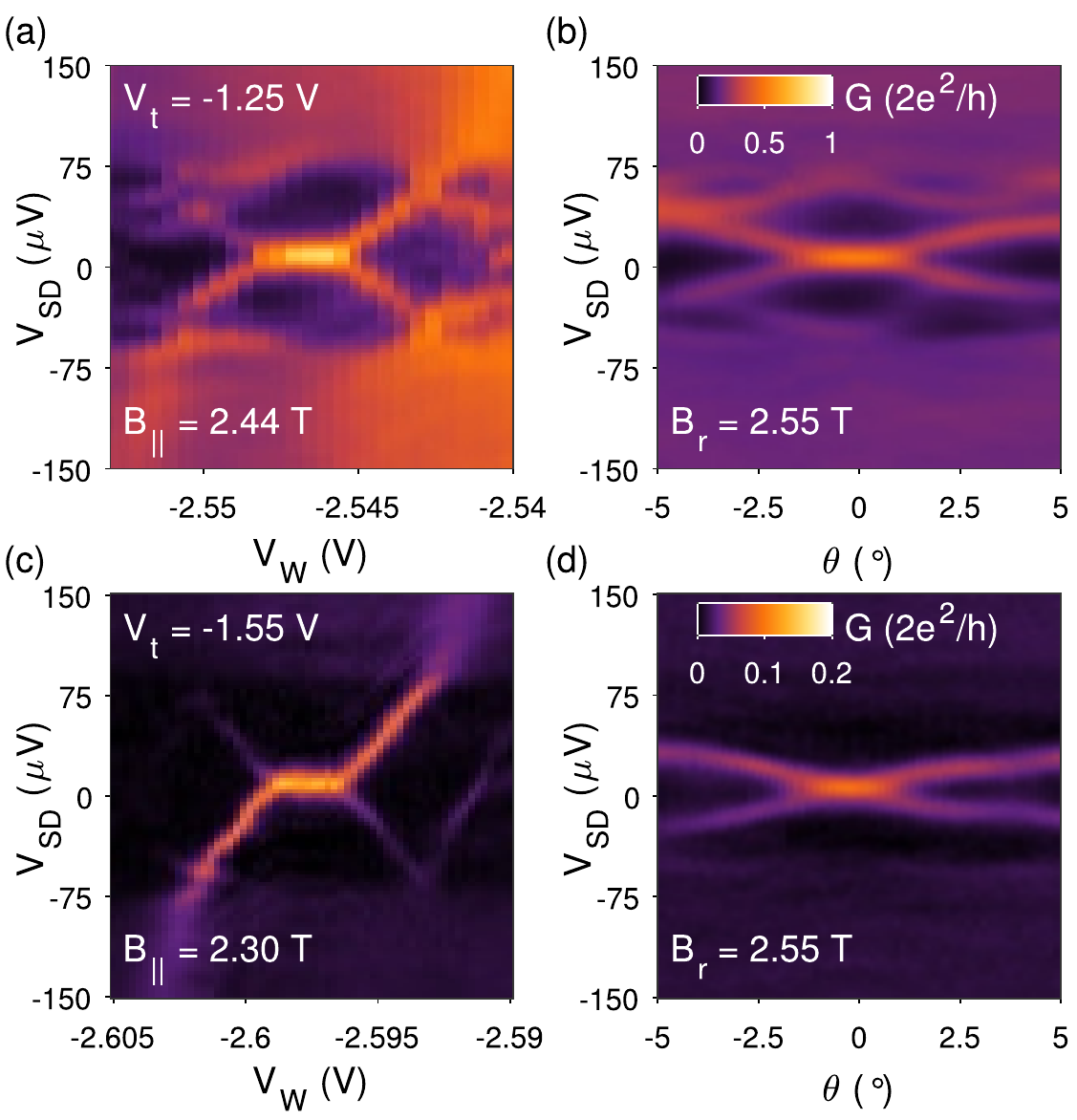}
\caption{(a) Tunneling spectroscopy as a function of $\Vw$ for $\Vt=-1.25~\rm{V}$. (b) Tunneling spectroscopy as a function of $\theta$ for $\Vt=-1.25~\rm{V}$. $\Vw$ is set in the center of the ZBP region of (a). (c,d) The same as in (a,b) for $\Vt=-1.55~\rm{V}$. Subfigures (a) and (b), as well as (c) and (d) share the same colorbar.}
\label{figs2}
\end{figure}

\begin{figure}
\includegraphics[width=\columnwidth]{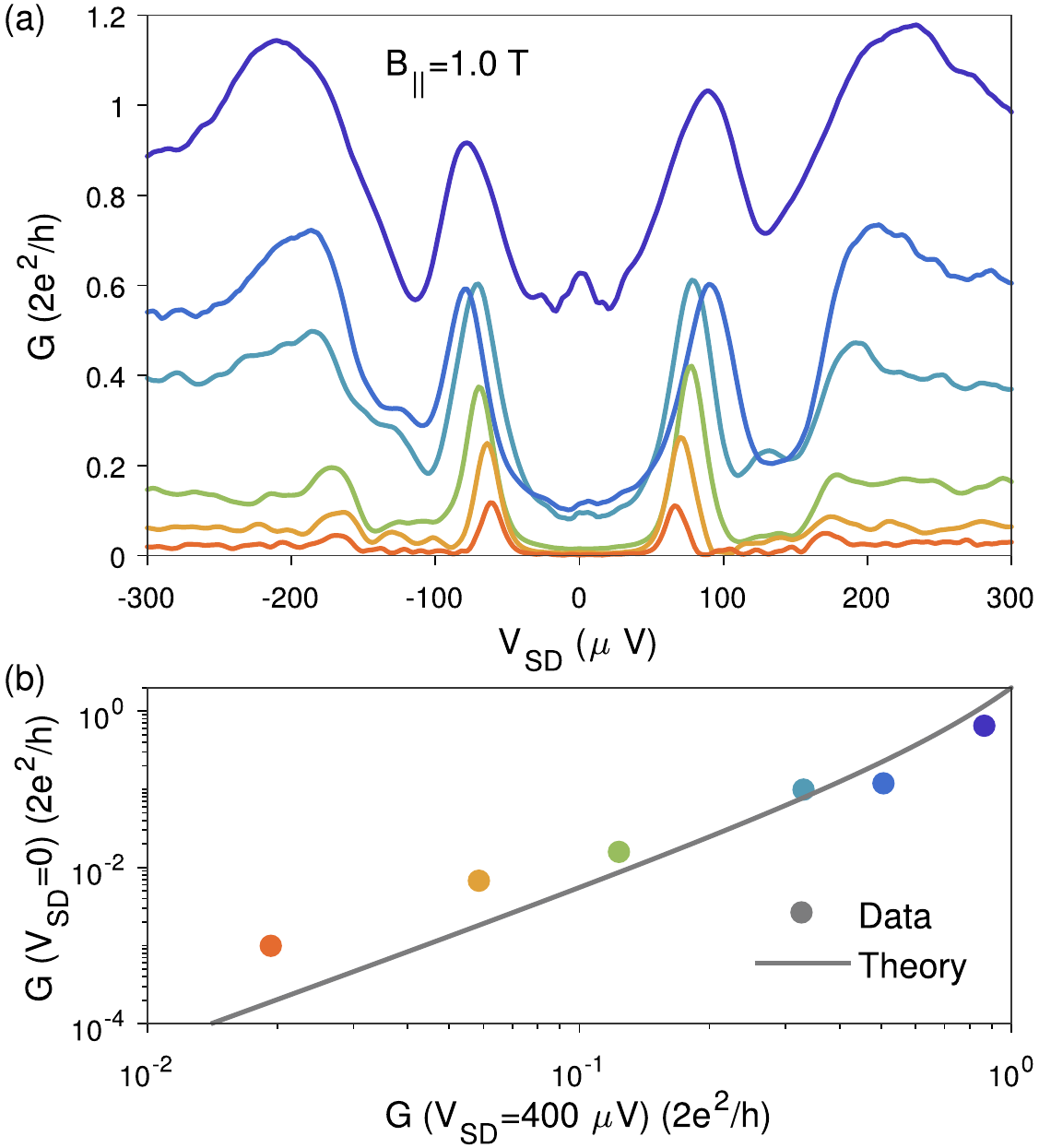}
\caption{Tunneling conductances for Sample~1 measured at $\Bp=1.0~\rm{T}$ for different values of normal state conductance $\Gn$. The colors are the same as in Fig.~\ref{fig1}(e) of the Main Text. Tunneling conductance for $\Vsd=0$ parametrically plotted versus the conductance at large $\Vsd$ together with a theoretical curve for a single mode superconductor/normal interface with no free parameters~\cite{Beenakker1992,Kjaergaard2017}.}
\label{figs3_5}
\end{figure}

Electrical measurements were performed with conventional DC and AC lock-in techniques using low frequency ($<200~\rm{Hz}$) excitations. An AC voltage bias of $3~\rm{\mu V}$, superimposed to a variable DC bias, was applied to one lead of the wire, with the other lead grounded via a low impedance current to voltage converter. An AC voltage amplifier with an input impedance of $500~\rm{M\Omega}$ measured the four terminal voltage across the strip. Throughout this manuscript, we will use the symbol $G$ to indicate the differential conductance $dI/dV$.

The samples were studied in a dilution refrigerator with a base temperature $T=20~\rm{mK}$, measured with a calibrated $\rm{RuO_x}$ thermometer mounted on the mixing chamber plate. The mixing chamber temperature was tuned with an electrical heater and a PID controller.

In our measurements, we saw the first temperature dependences of the conductance at $T\approx30~\rm{mK}$. Therefore, we present only data taken for $T\geq40~\rm{mK}$, where we assume the electron temperature and mixing chamber temperature are approximately equal. This assumption finds agreement to an independent temperature fit of the ZBP conductance of Fig.~\ref{fig2}(a) of the Main Text with $\Gn$ and $\Dt$ as fixed parameters.
Measurements such as Figs.~\ref{fig1}(c) and~\ref{fig1}(d) of the Main Text and Fig.~\ref{figs1}(f) are obtained by sweeping $\Vsd$ and stepping $\Bp$ at a rate of $100~\rm{mT min^{-1}}$, resulting in additional heating of the device due to eddy currents in the refrigerator, which in turn results in lower ZBPs. We therefore use the data in Fig.~\ref{fig2}(a) of the Main Text, where the magnetic field was kept stable for a long time, as a reference to determine the effective electron temperature during a magnetic field scan. The analysis yields an effective electron temperature of $55~\rm{mK}$ during scans as those in Figs.~\ref{fig1}(c) and~\ref{fig1}(d). We therefore use $T=55~\rm{mK}$ for calculating the theory curves presented in Fig.~\ref{fig3}(b) of the Main Text, which result in an exceptionally good agreement with theory without any fitting parameters. Using $T=40~\rm{mK}$ results in theory curves with the same trend, but with an overall increase of the conductance by $\sim30\%$.

\section{Additional Characterization of Sample 1}
We provide here additional characterization measurement of Sample~1, which was presented in the Main Text. In addition to the peaks stability in an in-plane magnetic field $\Bp$, oriented along the wire direction [evident from Figs.~\ref{fig1}(c) and~\ref{fig1}(d) of the Main Text], we show here the ZBP behavior as a function of gate voltages and magnetic field orientation. The junction behavior at intermediate fields is indicative of a single mode superconductor/normal interface.

Figures~\ref{figs1}(a-e) show tunneling spectroscopy as a function of $\Vw$ with $\Vt=-1.40~\rm{V}$, which sets the normal state conductance to $\Gn\approx0.12\Go$. In this measurements $\Bp$ gradually varies from $1~\rm{T}$ [Fig.~\ref{figs1}(e)] to $2.8~\rm{T}$ [Fig.~\ref{figs1}(a)].
Andreev bound states, emerging from the gap edge at low field evolve into an extended ZBP, robust against small variation of $\Vw$, as $\Bp$ increases. At the same time, the superconducting energy gap gradually collapses. The entire magnetic field dependence for $\Vw=-2.555~\rm{V}$ is presented in Fig.~\ref{figs1}(f), qualitatively similar to the data presented in the Main Text. As discussed in the Main Text and in Ref.~\onlinecite{Suominen2017} the $4\Delta^*\rightarrow 2\Delta^*$ transition at $\Bp\approx300~\rm{mT}$ coincides with the onset of a finite density of states within the superconducting gap of the planar Al region.

\begin{figure}
\includegraphics[width=\columnwidth]{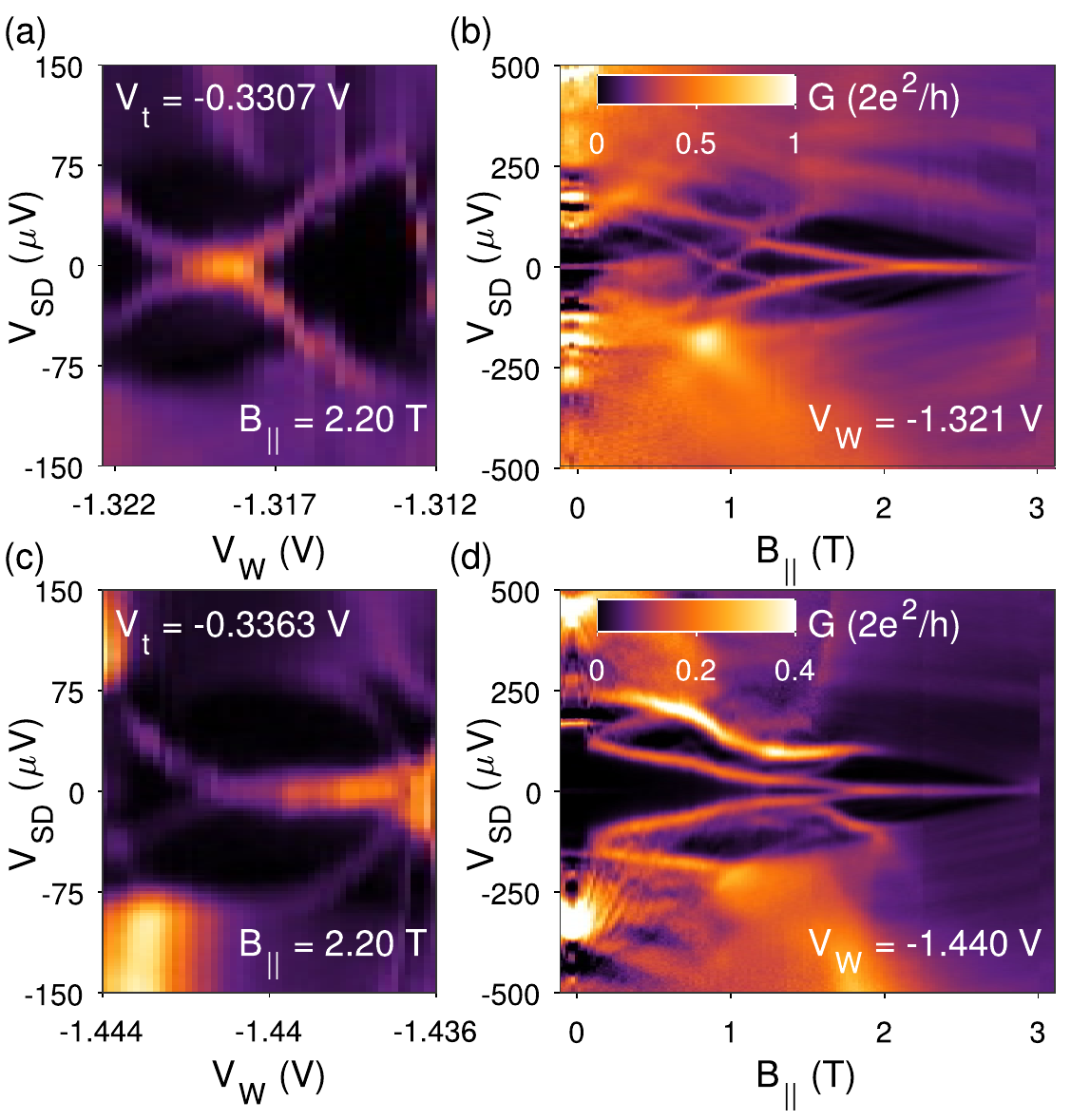}
\caption{Characterization of Sample 2. (a) Tunneling spectroscopy of ZBP1 as a function of $\Vw$ for $\Vt=-0.3307~\rm{V}$. (b) Tunneling spectroscopy as a function of $\Bp$ in the same configuration as (a) for $\Vw=-1.321~\rm{V}$. (c,d) Tunneling spectroscopy of ZBP2 as a function of $\Vw$ for $\Vt=-0.3363~\rm{V}$. (c) Tunneling spectroscopy as a function of $\Bp$ in the same configuration as (a) for $\Vw=-1.440~\rm{V}$. Subfigures (a) and (b), as well as (c) and (d) share the same colorbar.}
\label{figs3}
\end{figure}

In our setup, Majorana modes are predicted to form for large in-plane magnetic field $\Bp$ aligned perpendicular to the spin-orbit field, i.e. parallel to the Al strip. In Fig.~\ref{figs1}(g), we fix an in-plane magnetic field magnitude $\Br=2.55~\rm{T}$ and vary its orientation $\theta$ in the 2DEG plane, with $\theta=0$ being along the Al wire. For $|\theta|\leq4^\circ$, the ZBP is stable. For $|\theta|>4^\circ$, the peaks splits into two states that move to higher energy. The superconducting gap is not affected by the field orientation, indicating the Al strip persists in the superconducting regime regardless of the in-plane field orientation. Figure~\ref{figs1}(h) shows a similar measurement, where the field orientation $\phi$ was varied in a plane parallel the Al wire and perpendicular to the 2DEG. In this situation, the external field is always perpendicular to the spin-orbit direction so that the topological criterion is always met (provided a high enough Zeeman energy). As expected, a robust ZBP is observed as long as the superconducting gap persists. The quenching of the peak for $|\phi|>5^\circ$ is due to the large out-of-plane component, which drives the Al strip normal.

Figure~\ref{figs2} presents further characterization of the same ZBP performed with different values of $\Vt$, confirming an almost identical behavior for a broad range of transmissions. Figures.~\ref{figs2}(a) and (b) show the tunneling conductance as a function of $\Vw$ and $\theta$ performed for $\Vt=-1.25~\rm{V}$, setting $\Gn=0.23\Go$. Figures~\ref{figs2}(c) and (d) show the same analysis for $\Vt=-1.55~\rm{V}$, setting $\Gn=0.019\Go$ [as in Fig.~1(d) of the Main Text]. In both cases, the stability of the ZBP over a finite range of tuning parameters is confirmed.

The stability of the ZBP in a finite range of gate voltage and magnetic field is an expected feature of MZMs. A ZBP that was robust as a function of magnetic field, superficially resembling the measurement of Fig.~\ref{figs1}(f), was reported in superconductor/semiconductor devices at the crossing of Andreev states \cite{Lee2013}. A key difference between our observations and Ref. \onlinecite{Lee2013} is the extended gate voltage range in which the ZBP is observed. The ZBP extends for a $\Vw$ range of approximately $4~\rm{mV}$, which can be converted to a change in chemical potential of $96~\rm{\mu eV}$ using a wire gate lever arm of $24~\rm{meV/V}$ [extracted from the slope of the Andreev states visible in Fig.~\ref{figs2}(c)]. The energy extent of the ZBP is thus approximately six times larger than its FWHM [see Fig.~\ref{figs6}], ruling out an accidental crossing of trivial Andreev states. The energy span of $96~\rm{\mu eV}$ might serve as a measurement for the helical gap opening in the band structure of the wire. Given a $g$-factor of $4$ and a magnetic field of $2.3~\rm{T}$, a helical gap of $g\mu_{\rm{B}}B\approx500~\rm{\mu eV}$ is expected for strong spin-orbit coupling. The gate extent of the observed ZBP is therefore reasonable, given that multi-mode transport and disorder presumably reduce the size of the helical gap.

We conclude the analysis of Sample 1 by studying the quality of its tunneling probe in the superconductor/normal configuration. We perform the analysis from tunneling conductance curves for various transmission values taken at $\Bp=1~\rm{T}$ [Fig.~\ref{figs3_5}(a)], where the supercurrent is largely suppressed and the high field ZBP did not form yet. Following the method of Ref.~\onlinecite{Kjaergaard2017}, we plot the in-gap conductance (measured for $\Vsd=0$) against the conductance measured in the same configuration but for large $\Vsd$. The experimental points are shown in Fig.~\ref{figs3_5}(b) (markers) and compared to a theory for a single mode ballistic superconductor/normal interface \cite{Beenakker1992} (solid curve). The reasonably good agreement over a large ranges of normal state conductances, compatible with our previous studies performed at zero magnetic field \cite{Kjaergaard2017}, demonstrate the tunneling probe is single mode and ballistic.

\section{Characterization of Sample 2}
Similar measurements were performed on Sample~2. Differently from Sample~1, in Sample~2 we could not identify ZBPs that were robust against a 3 orders of magnitude variation of $\Gn$. Therefore, we conducted the analysis on two distinct ZBPs, named ZBP1 and ZBP2, obtained for similar values of $\Vw$. Each ZBP was robust for a moderate variation of $\Gn$ so that the overall range of transmission investigated for Sample~2 was comparable to that of Sample~1.

\begin{figure}
\includegraphics[width=\columnwidth]{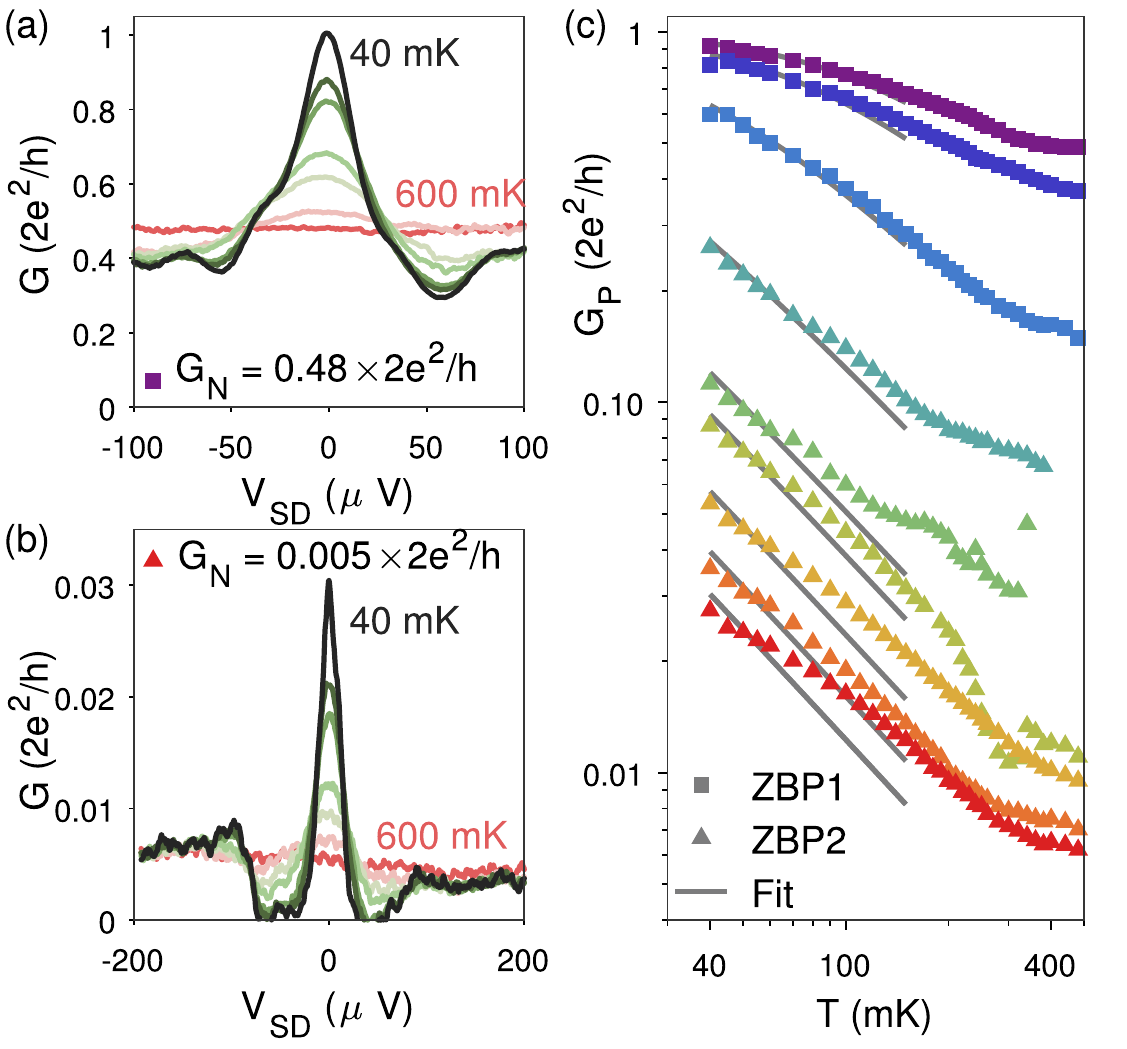}
\caption{(a) Temperature dependence of ZBP1 at $\Bp =2.2~\rm{T}$ and $\Gn=0.48\Go$. Temperatures are $40$ (black), $60$, $80$, $100$, $150$, $200$, $300$ and $600~\rm{mK}$ (pink). As (a) but for ZBP2 with $\Gn=0.005\Go$. (c) Temperature dependence of $\Gp$ for ZBP1 (squares) and ZBP2 (triangles) as a function of $\Gn$. Values of $Gn$ range from $0.48$ (top) to $0.005$ (bottom).}
\label{figs4}
\end{figure}

Measurements of ZBP1, characterized by a large transmission, are presented in Figs.~\ref{figs3}(a) and (b) while those of ZBP2, characterized by a low transmission, are presented in Figs.~\ref{figs3}(c) and (d). Both ZBPs arise from coalescing discrete states at $\Bp\approx 2~\rm{T}$, and are robust against moderate variation of $\Vw$. In Sample~2, a larger number of states populate the gap at low magnetic fields, compared to Sample~1. Once the ZBPs form, the trivial levels have already moved outside the superconducting energy gap, not precluding our study. The rehardening of the induced gap at high magnetic field might also be an indication of the reopening of the superconducting gap after the topological transition.

\begin{figure}
\includegraphics[width=\columnwidth]{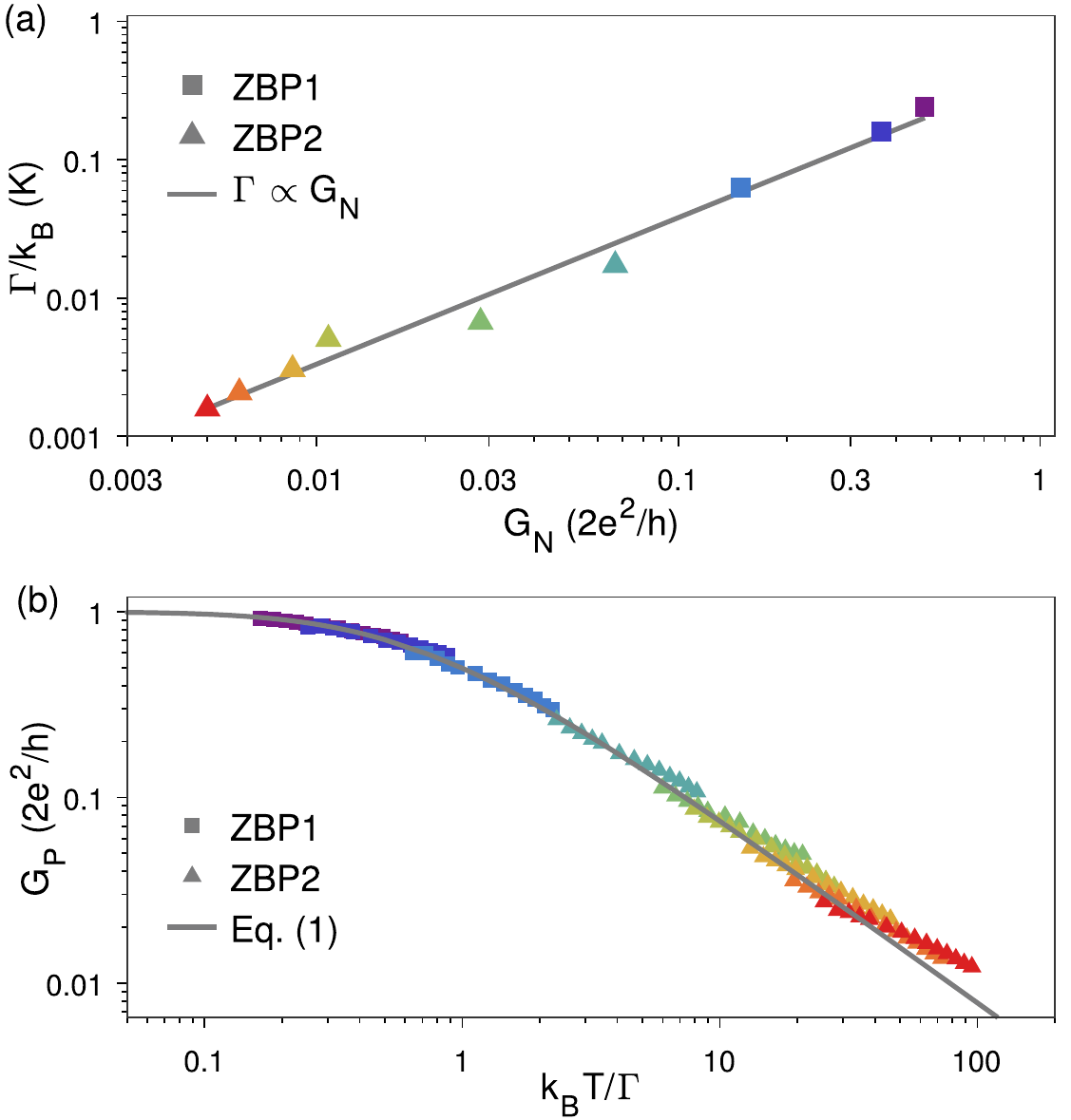}
\caption{(a) Extracted relation between the energy broadening $\Gamma$ obtained from the fits in Fig.~\ref{figs4}(c) and the normal state conductance $\Gn$ (squares and triangles for ZBP1 and ZBP2 respectively). The solid line is a linear fit, which reveals proportionality between $\Gamma$ and $\Gn$. (b) Entire dataset (for $T\leq150~\rm{mK}$) plotted as a function of temperature scaled by the extrated values of $\Gamma$ (squares and triangles for ZBP1 and ZBP2 respectively) and compared to the theoretical expectation (solid line).}
\label{figs5}
\end{figure}

Exemplary temperature dependences of ZBP1 and ZBP2 are shown in Figs.~\ref{figs4}(a) and (b) respectively. The low temperature conductance of ZBP1 at $\Vsd=0$ is very close to $2e^2/h$. The entire set of ZBP heights $\Gp$ as a function of temperature is presented in Figs.~\ref{figs4}(c), with squares for ZBP1 and triangles for ZBP2, respectively. As for Sample~1, for low transmission $\Gp$ varies steeply as a function of temperature. On the contrary, the temperature dependence is quenched for large $\Gn$, with $\Gp\approx2e^2/h$ independent of transmission. The gray lines are fit to Eq.~\ref{Glowtemp} of the Main Text.

Following the same analysis presented in the Main Text, in Fig.~\ref{figs5}(a) we plot the relationship between the values of $\Gamma$ extracted from the fit of Fig.~\ref{figs4}(c) and $\Gn$ measured from the high bias differential conductance. Also in this case, a power law fit reveals a linear relationship. The solid curve in Fig.~\ref{figs5}(a) is a linear fit, giving a proportionality constant of $360~\rm{mK}/(2e^2/h)$, similar to Sample~1.
The data in Fig.~\ref{figs4}(c) can then be plotted as a function of the scaling parameter $k_{\rm{B}}T/\Gamma$ [Fig.~\ref{figs5}(b)], reproducing the theoretical prediction of Eq.~\ref{Glowtemp} of the Main Text over 3 orders of magnitude.

\section{Width of the Zero Bias Peaks}
\begin{figure}
\includegraphics[width=\columnwidth]{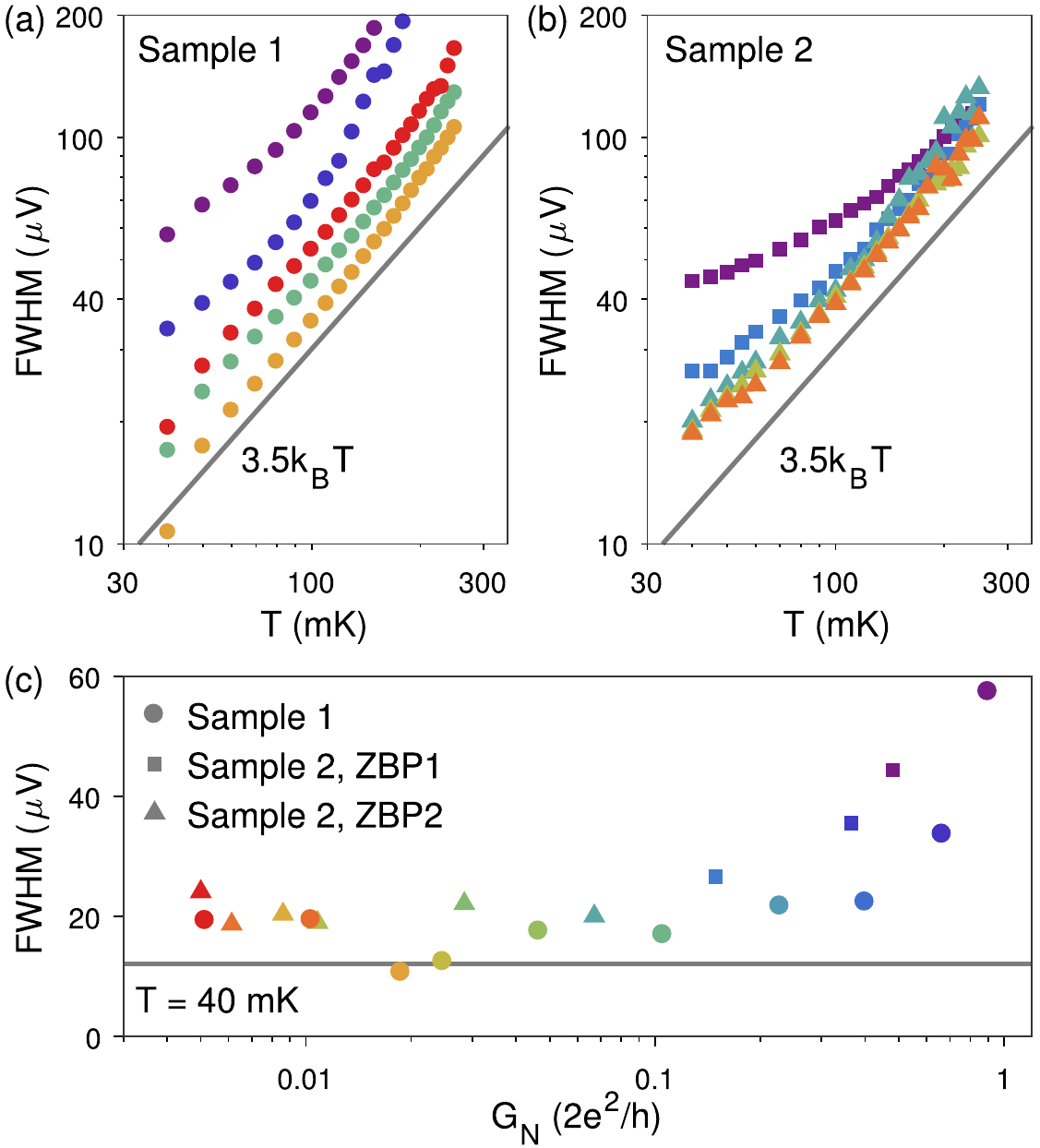}
\caption{(a) FWHM as a function of temperature for Sample 1. Colors refer to normal state conductance values, with the same color coding as Fig.~\ref{fig1} of the Main Text. (b) FWHM as a function of temperature for Sample 2. Colors refer to normal state conductance values, with the same color coding as Fig.~\ref{figs4}. (c) FWHM of both samples measured at the fridge base temperature. The gray line indicates the expected Lorentzian linewidth for $T=40~\rm{mK}$.}
\label{figs6}
\end{figure}

Figures~\ref{figs6}(a,b) show the FWHM of the ZBPs measured in Sample~1 and Sample~2, as a function of temperature, for different values of $\Gn$ and for $\Bp=2.3~\rm{T}$. The FWHM is extracted by fitting the data to a Lorentzian function, with no background subtracted. The color coding is the same as Fig.~\ref{fig2}(c) of the Main Text and Fig.~\ref{figs4}(c) for Fig.~\ref{figs6}(a) and (b) respectively. The solid line represents the relation $\rm{FWHM}=3.5k_{\rm{B}}T$ expected in case of a Lorentzian function broadened by temperature only. Figure~\ref{figs6}(c) shows the $T=40~\rm{mK}$ FWHM as a function of normal state conductance for both samples, with the gray line being the expected FWHM for temperature dominated broadening ($12~\rm{\mu eV}$). We note that the large width ($\rm{FWWHM}\sim\Delta^*$) of the ZBPs for high transmission renders the Lorentzian fits inaccurate, especially at the high temperature, where the background contribution is expected to be significant. 

\end{document}